\documentclass{elsart5p}
\usepackage{graphics}
\usepackage{graphicx}
\usepackage{epsfig}
\usepackage{amssymb}
\usepackage{amsmath}

\def\fmn#1#2{\mbox{${\textstyle \frac{#1}{#2}}$}}
\newcommand{\pppi}{\mbox{$pp\to \{pp\}_{\!s\,}\pi^0$}}
\newcommand{\vpppi}{\mbox{$\pol{p}p\to \{pp\}_{\!s\,}\pi^0$}}
\newcommand{\nppppi}{\mbox{$np\to \{pp\}_s\pi^-$}}
\begin{document}

\begin{frontmatter}

\title{Differential cross section and analysing power of the
$pp\to\{pp\}_s\pi^0$ reaction at 353~MeV}

\author[dubna]{D.~Tsirkov},
\author[dubna]{T.~Azaryan},
\author[bochum,itep]{V.~Baru},
\author[ikp]{D.~Chiladze},
\author[erlangen,dubna]{S.~Dymov},
\author[gatchina]{A.~Dzyuba},
\author[ikp]{R.~Gebel},
\author[munster]{P.~Goslawski},
\author[ikp]{C.~Hanhart},
\author[ikp]{M.~Hartmann},
\author[ikp]{A.~Kacharava},
\author[munster]{A.~Khoukaz},
\author[dubna]{V.~Komarov},
\author[cracow]{P.~Kulessa},
\author[dubna]{A.~Kulikov},
\author[dubna]{V.~Kurbatov},
\author[manchester,itep]{V.~Lensky},
\author[ikp]{B.~Lorentz},
\author[dubna,tbilisi]{G.~Macharashvili},
\author[tbilisi]{D.~Mchedslishvili},
\author[munster]{M.~Mielke},
\author[gatchina]{S.~Mikirtytchiants},
\author[dubna,ikp]{S.~Merzliakov},
\author[tbilisi]{M.~Nioradze},
\author[ikp]{H.~Ohm},
\author[munster]{M.~Papenbrock},
\author[ikp]{F.~Rathmann},
\author[dubna,ikp]{V.~Serdyuk},
\author[dubna]{V.~Shmakova},
\author[ikp]{H.~Str\"oher},
\author[dubna]{Yu.~Uzikov},
\author[gatchina,ikp]{Yu.~Valdau},
\author[ucl]{C.~Wilkin\corauthref{cor1}}
\ead{cw@hep.ucl.ac.uk} \corauth[cor1]{Corresponding author.}

\address[dubna]{Laboratory of Nuclear Problems, Joint Institute for Nuclear
  Research, RU-141980 Dubna, Russia}
\address[bochum]{Institut f\"ur Theoretische Physik II, Ruhr-Universit\"at Bochum, D-44780 Bochum, Germany}
\address[itep]{Institute for Theoretical and Experimental Physics, RU-117218 Moscow,
Russia}
\address[ikp]{Institut f\"ur Kernphysik, Forschungszentrum J\"ulich, D-52425
  J\"ulich, Germany}
\address[erlangen]{Physikalisches Institut II, Universit{\"a}t
Erlangen-N{\"u}rnberg, D-91058 Erlangen, Germany }
\address[gatchina]{St. Petersburg Nuclear Physics Institute, RU-188350 Gatchina,
  Russia}
\address[munster]{Institut f\"ur Kernphysik, Universit\"at M\"unster,
D-48149 M\"unster, Germany}
\address[cracow]{Institute of Nuclear Physics, PL-31342 Cracow,
Poland}
\address[tbilisi]{High Energy Physics Institute, Tbilisi State University, GE-0186
Tbilisi, Georgia}
\address[manchester]{School of Physics and Astronomy, University of Manchester, Manchester M13 9PL,
UK}
\address[ucl]{Physics and Astronomy Department, UCL, London WC1E 6BT, UK}

\begin{abstract}
In order to establish links between $p$-wave pion production in
nucleon-nucleon collisions and low energy three-nucleon scattering, an
extensive programme of experiments on pion production is currently underway
at COSY-ANKE. The final proton pair is detected at very low excitation
energy, leading to an $S$-wave diproton, denoted here as $\{pp\}_{\!s}$. We
now report on measurements of the differential cross section and analysing
power of the $\pol{p}p\to\{pp\}_{\!s\,}\pi^0$ reaction at 353~MeV. Both observables
can be described in terms of $s$- and $d$-wave pion production and, by using
the phase information from elastic $pp$ scattering, unique solutions can be
obtained for the corresponding amplitudes. This information is vital for the
partial wave decomposition of the corresponding $pn\to\{pp\}_{\!s\,}\pi^-$
reaction and hence for the extraction of the $p$-wave terms.
\end{abstract}

\end{frontmatter}


Within the context of chiral perturbation theory, a significant
step forward in our understanding of pion physics at low
energies would be to establish that the same short-ranged
$NN\to NN\pi$ vertex contributes to $p$-wave pion production,
to low energy three-nucleon
scattering~\cite{HAN2000,EPE2002,BAR2009}, $\gamma d\to\pi
NN$~\cite{LEN2005,LEN2007} and $\pi d\to \gamma
NN$~\cite{GAR2006}, as well as in weak reactions like tritium
beta decay~\cite{PAR2003,GAR2006a,NAK2008,GAZ2009}. The
relevant transition amplitude, which connects $NN$ $S$-waves in
the initial and final state with a $p$-wave pion, contributes
to both $pp\to \pi^+d(\pi^+pn)$ and $pn\to pp\pi^-$. However,
the extensive data for $\pi^+$ production is of limited use in
this context, because the $p$-wave amplitudes are completely
dominated by the $^{1\!}D_2$ initial state, which hinders a
reliable extraction of the $^{1\!}S_0$ initial
state~\cite{BAR2009}.

There is a programme at the COSY-ANKE facility of the Forschungszentrum
J\"ulich to perform a complete set of measurements on
$NN\to\{pp\}_{\!s\,}\pi$ at low energy~\cite{SPIN}. Here the $\{pp\}_{\!s}$
denotes a proton-proton system with very low excitation energy, $E_{pp}$. At
ANKE we select events with $E_{pp}<3$~MeV and, under these conditions, the
diproton is overwhelmingly in the $^{1\!}S_0$ state with antiparallel proton
spins. This simplifies enormously the spin structure: a partial wave analysis
for $pp\to pp\pi^0$ without the $E_{pp}$ cut would require twelve additional
$P$-wave final $pp$ spin-triplet states~\cite{MEY1999,DEE2005,HAN2004}. The
cut also allows one to extract the full information on the production
amplitudes without having to make measurements of the final proton
polarisations.

Whereas the $\{pp\}_{\!s}$ final state is isotriplet, the isosinglet $np$
initial state also contributes to $pn\to pp\pi^-$. In order to isolate this,
which contains the amplitudes of interest, the isotriplet channel needs to be
well understood. As the first part of the outlined larger programme, we
therefore report here on measurements of the cross section and proton
analysing power in the \pppi\ reaction at $T_p=353$~MeV.

For a spin-singlet diproton, the spin structure of the \pppi\ or \nppppi\
reaction is that of $\half^+\half^+\to 0^+0^-$. Parity and angular momentum
conservation require that the initial nucleon-nucleon pair to have spin
$S=1$. The pion orbital angular momentum $\ell$ and the initial
nucleon-nucleon isospin $I$ are then linked by $\ell + I =$~odd so that, for
the \pppi\ reaction, only even pion partial waves are allowed. As a
consequence, the unpolarised cross section for $\pi^0$ production, and this
times the proton analysing power $A_y$, must be of the form
\begin{eqnarray}
\label{dsig}
\left(\frac{d\sigma}{d\Omega}\right)_{\!0}&=&\frac{k}{4p}\left(a_0
+a_2\cos^2\theta_{\pi} + a_4\cos^4\theta_{\pi} +\cdots\right),\\
\label{aydsig}
A_y\left(\frac{d\sigma}{d\Omega}\right)_{\!0}&=&\frac{k}{4p}
\sin\theta_{\pi}\cos\theta_{\pi}\left(b_2 +b_4\cos^2\theta_{\pi} +\cdots\right),
\end{eqnarray}
where $\theta_{\pi}$ is the pion c.m.\ production angle with respect to the
direction of the polarised proton beam. Here $p$ is the incident c.m.\
momentum and $k$ that of the produced pion which, at 353~MeV, have values
$p=407$~MeV/$c$ and $k\approx 94$~MeV/$c$, where the latter represents an
average over the 3~MeV $E_{pp}$ range.

The only detailed measurements of the \pppi\ differential cross section over
the whole angular range were carried out with the PROMICE-WASA apparatus at
CELSIUS at a series of energies from 310 to 450~MeV, using the same standard
3~MeV cut on $E_{pp}$~\cite{BIL2001}. Throughout this energy range,
significant anisotropies were found in the angular distributions which were
attributed to interferences between pion $s$ and $d$ waves. On the other
hand, there were no corresponding measurements of the proton analysing power,
which might also be driven by a strong $s$-$d$ interference.

We have previously reported measurements of the \pppi\ differential cross
section at several energies and small angles~\cite{DYM2006,KUR2008}. Since
these were carried out using the ANKE spectrometer~\cite{BAR1997} under
conditions that were similar to the current ones, the description here can be
quite brief. ANKE is placed at an internal beam station of the COSY cooler
synchrotron. Fast charged particles, resulting from the interaction of the
stored transversally polarised proton beam with the hydrogen cluster-jet
target~\cite{KHO1999} and passing through the analysing magnetic field, were
recorded in the forward detector (FD) system. The FD, which was the only
detector used in this experiment, includes multiwire proportional chambers
for tracking and a scintillation counter hodoscope for energy loss and timing
measurements.

To start the identification of the \pppi\ reaction, proton pairs were first
selected from all the registered two-track events using the measured momenta
of the both particles and the difference in their
time-of-flight~\cite{CYR2010}. The resolution $\sigma(E_{pp})$ in the
diproton excitation energy was better than 0.6~MeV, which allowed the
$E_{pp}<3$~MeV cut to be applied reliably.

\begin{figure}[ht]
\centering
\includegraphics[width=0.99\columnwidth]{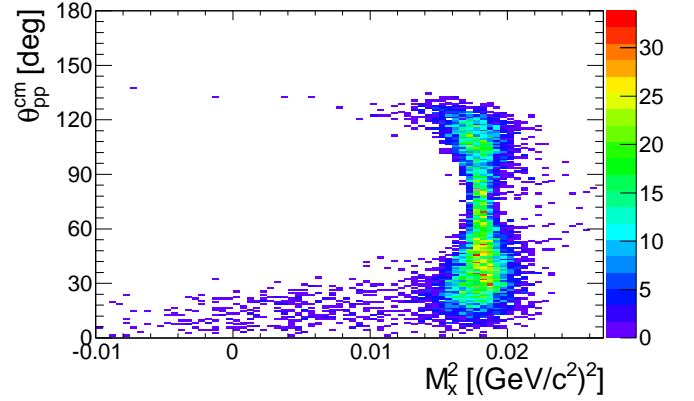}
\caption{\label{gig1} Two-dimensional distribution of the
missing-mass-squared $M_{\!X}^2$ of the $pp\to\{pp\}_{\!s}X$ reaction
at 353~MeV versus the diproton c.m.\ polar angle
$\theta_{pp}^\mathrm{cm}$ for events with $E_{pp}<3$~MeV.}
\end{figure}

After selecting the $^{1\!}S_0$ final state, the kinematics of the
$pp\to\{pp\}_{\!s}X$ process could be reconstructed on an event-by-event
basis to obtain a missing-mass $M_{\!X}$ spectrum. A two-dimensional
distribution of $M_{\!X}^2$ versus the c.m.\ polar angle of the diproton
$\theta_{pp}^\mathrm{cm}$ is presented in Fig.~\ref{gig1}. This demonstrates
the large angular acceptance of the apparatus for the \pppi\ reaction at
353~MeV and shows a clean $\pi^0$ signal with an almost negligible
background. Simulations indicate that the c.m.\ angular resolution is better
than $5^\circ$.

The polarization asymmetry is defined by
\begin{equation}
\label{epsilon}
\varepsilon=\frac{N_\uparrow/L_\uparrow-N_\downarrow/L_\downarrow}
{N_\uparrow/L_\uparrow+N_\downarrow/L_\downarrow},
\end{equation}
where $N_\uparrow$ and $N_\downarrow$ are the numbers of \pppi\ events with
beam proton spin up and down, corrected for dead time, and $L_\uparrow$ and
$L_\downarrow$ are the corresponding luminosities. The relative luminosity
$L_{\uparrow}/L_{\downarrow}\approx 0.985 \pm 0.015$ was estimated using
events at very small polar angles, where the polarization asymmetry should
be negligible. This procedure adds about a 3\% systematic error to the
values of $\varepsilon$.

The analysing power $A_y$ is connected to the asymmetry
through:
\begin{equation}\label{Aydef}
A_y=\frac{\varepsilon}{P\,\langle\cos\phi_{pp}\rangle},
\end{equation}
where $P$ is the transverse polarization of the beam and
$\langle\cos\phi_{pp}\rangle$ the average over the diproton azimuthal angular
distribution. Since the $\cos\phi_{pp}$ acceptance is concentrated near 1,
all the events in the regions analysed contribute usefully to the $A_y$
measurement.

\begin{figure}[htb]
\centering
\includegraphics[width=0.9\columnwidth]{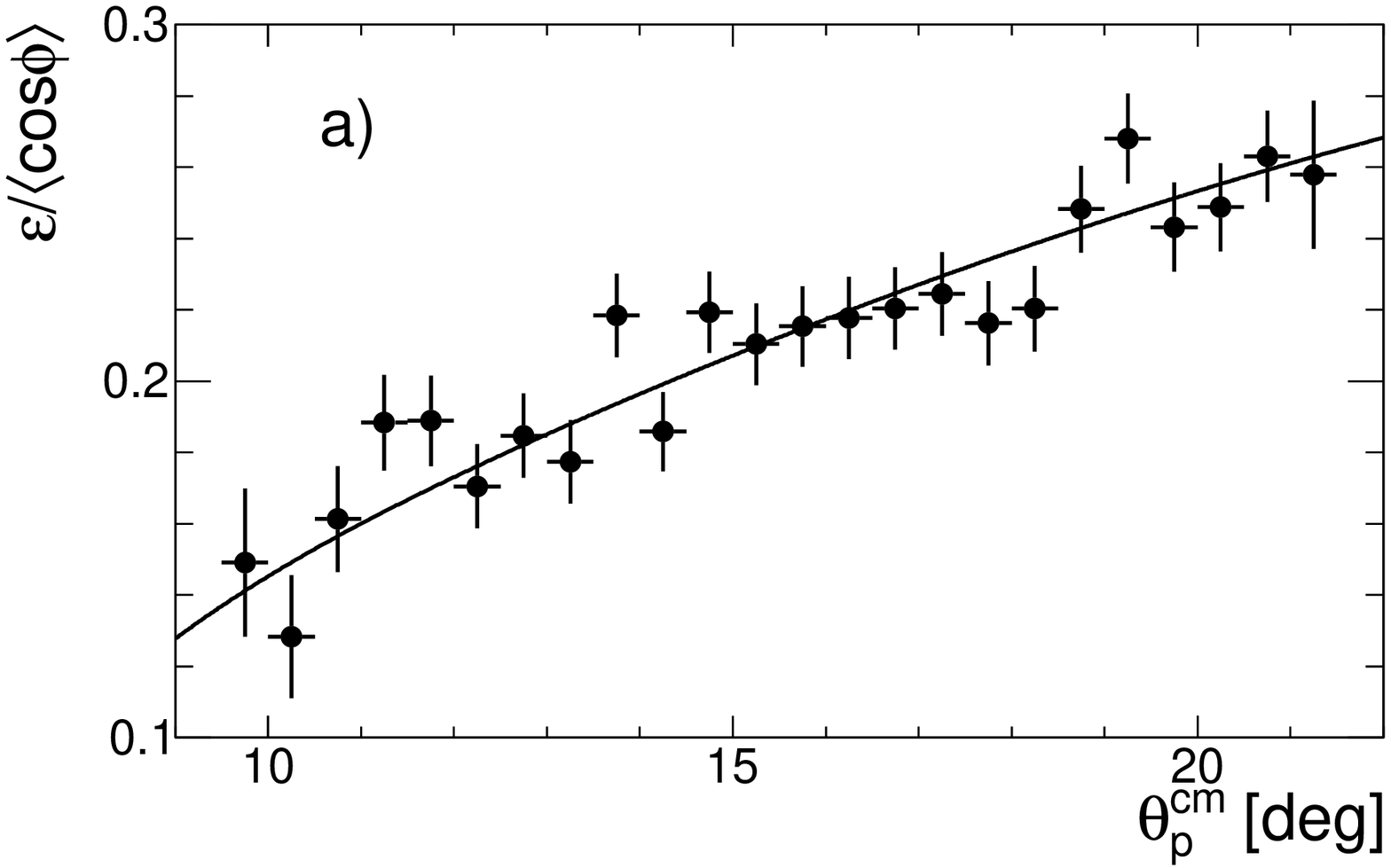}
\includegraphics[width=0.9\columnwidth]{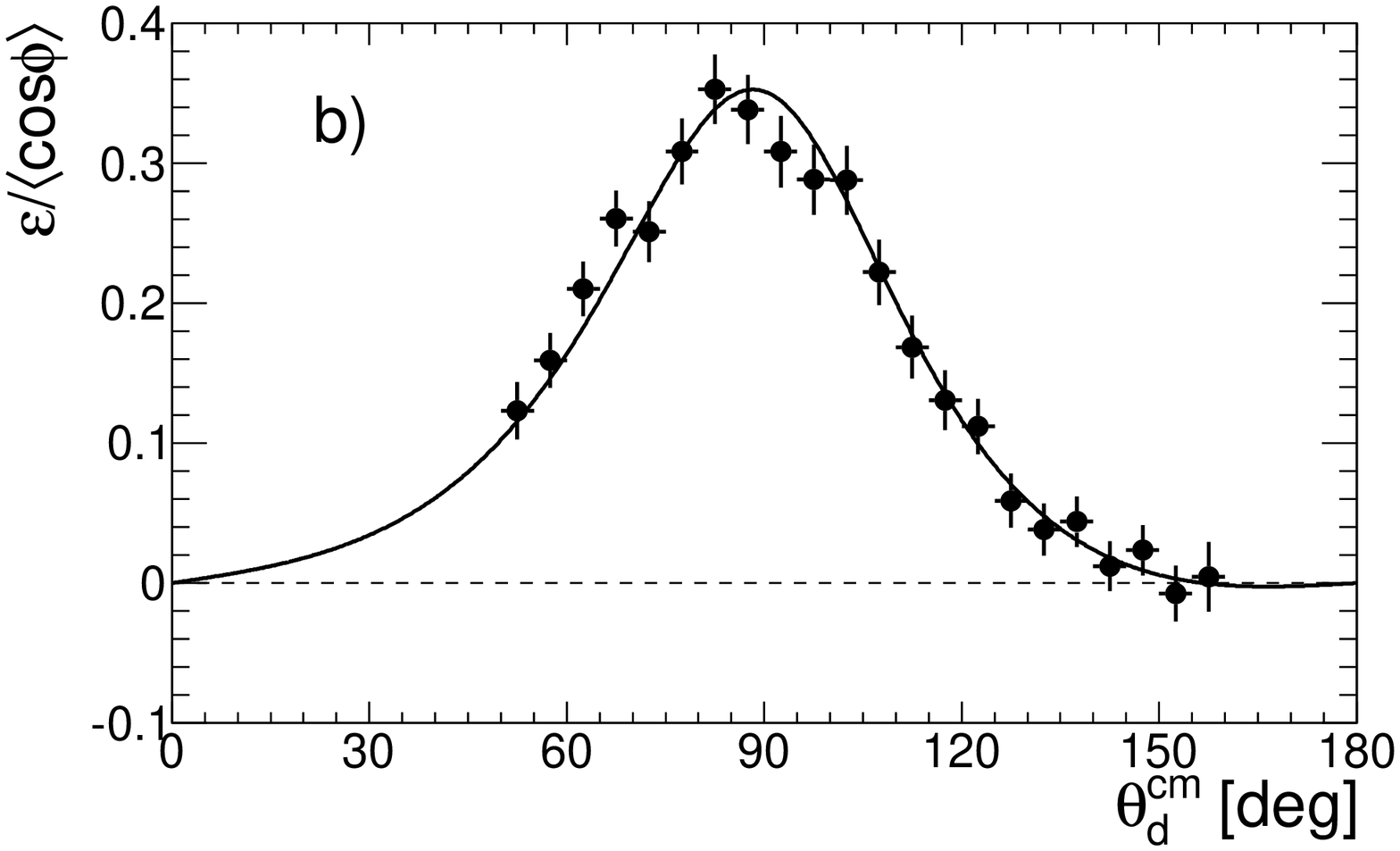}
\caption{\label{gig2} Product of the beam polarisation and analysing power at
a beam energy of 353~MeV for (a) elastic $pp$ scattering and (b) the $pp\to
d\pi^+$ reaction. The predictions of the SAID program~\cite{SAID} have been
scaled to agree with the experimental data and these give average COSY proton
beam polarisations of (a) $P = 0.687\pm0.008$ and (b) $P = 0.668\pm0.016$. In
neither case was the uncertainty in the SAID prediction included.}
\end{figure}

The polarization of the proton beam was flipped between ``spin-up'' to
``spin-down'' (perpendicular to the plane of the accelerator) every six
minutes and no measurements were made with an unpolarized beam. The value of
$P$ was estimated from proton-proton elastic scattering and the $pp\to
d\pi^+$ reaction that were measured in parallel. The analysing powers for
these reactions were taken from the SAID analysis program, solutions SP07 for
$pp\to pp$ and SP96 for $pp\to d\pi^+$~\cite{SAID}. The results of the two
methods shown in Fig.~\ref{gig2} agreed within measurement errors and gave an
average polarization of $P=0.68\pm0.03$, where the error includes the
uncertainties arising from the calibration reactions.

A simulation was undertaken of the two-dimensional acceptance in terms of the
$pp$ excitation energy $E_{pp}$ and its c.m.\ polar angle $\theta_{pp}$. This
took into account the geometry of the setup and the sensitive areas of the
detectors, the efficiency of the multiwire proportional chambers and the
track reconstruction algorithm. In order to avoid potential problems arising
near the limits of the acceptance, cuts were made around the edges of the
exit window of the spectrometer magnet in both the experimental data and
simulation. This is only a challenge at the larger angles,
$80^{\circ}<\theta_{\pi}<100^{\circ}$, where a compromise had to be made
regarding the acceptance ambiguities and this introduces an extra 4\%
systematic uncertainty in this angular region.

The numbers of detected $\pi^0$ events were then corrected on an
event-by-event basis for acceptance, dead time and relative luminosity
$L_{\uparrow}/L_{\downarrow}$. The latter were important because, in the
absence of data with an unpolarised beam, an average has to be evaluated.

The luminosity in the experiment was estimated from measurements of $pp$
elastic scattering carried out in parallel. The numbers of detected events,
corrected for the dead time, were compared with a simulation that used a
generator which included the differential cross section obtained from the
SAID analysis program~\cite{SAID}. Although this program does not furnish
error bars, experimental data at nearby energies suggests that the associated
uncertainty is about 2\%, to which must be added 3\% arising from acceptance
and similar systematic effects. At this level the statistical error is
negligible and the resulting total luminosity was estimated to be $544\pm
22$~nb$^{-1}$. At this energy the $pp\to d\pi^+$ cross section data are less
precise than those of $pp$ elastic scattering but, on the basis of the SAID
predictions, one obtains the completely consistent luminosity estimate of
547~nb$^{-1}$.

\begin{figure}[!ht]
\centering
\includegraphics[width=0.99\columnwidth]{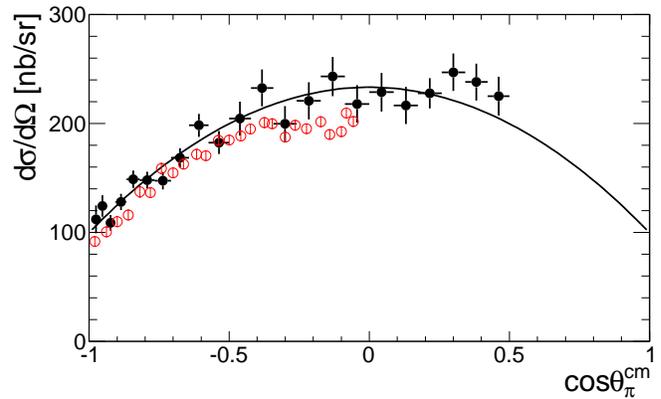}
\caption{\label{gig3} Differential cross section for the \pppi\ reaction at
353~MeV as a function of the cosine of the pion c.m.\ angle. The solid
(black) circles represent the ANKE measurements. The errors shown here are
statistical together with a 4\% systematic contribution in the
$80^{\circ}<\theta_{\pi}<100^{\circ}$ region coming from the acceptance
ambiguity discussed in the text. The overall systematic uncertainty is about
4\%. Open (red) circles are CELSIUS data obtained at 360~MeV~\cite{BIL2001}.
It should be noted that the latter data represent averages of measurements
taken in both hemispheres. The curve is a linear fit in $\cos^2\theta_{\pi}$
to our data. }
\end{figure}

The differential cross section results are presented in Fig.~\ref{gig3},
where they are compared to those obtained at 360~MeV at
CELSIUS~\cite{BIL2001}. Within the 10\% luminosity uncertainty in these data,
the overall agreement is very good. However, the CELSIUS data at this energy
level off a little around $90^\circ$. This seems to be a feature only of the
360~MeV results since, at the other energies, linear fits in
$\cos^2\theta_{\pi}$ all have good values of $\chi^2$/NDF~\cite{BIL2001}.

Fitting our data with a polynomial in $\cos^2\theta_{\pi}$, as in
Eq.~(\ref{dsig}), gives parameters
\begin{eqnarray}
\nonumber a_0&=&\phantom{-}4.05\pm0.08~\mu\textrm{b/sr},\\
a_2&=&-2.31\pm0.14~\mu\textrm{b/sr}, \label{sigval}
\end{eqnarray}
Apart from the acceptance uncertainties at the larger angles, the error bars
quoted here are purely statistical; the $\pm4\%$ systematic uncertainty from
the luminosity and acceptance largely cancels in the ratio $a_2/a_0$. Since
$\chi^2/\textrm{NDF}=23/20$, there is clearly no compelling evidence for any
$\cos^4\theta_{\pi}$ dependence, i.e., a non-zero $a_4$ coefficient, and this
possibility has been omitted from the curve in Fig.~\ref{gig3}.

\begin{figure}[!ht]
\centering
\includegraphics[width=0.95\columnwidth]{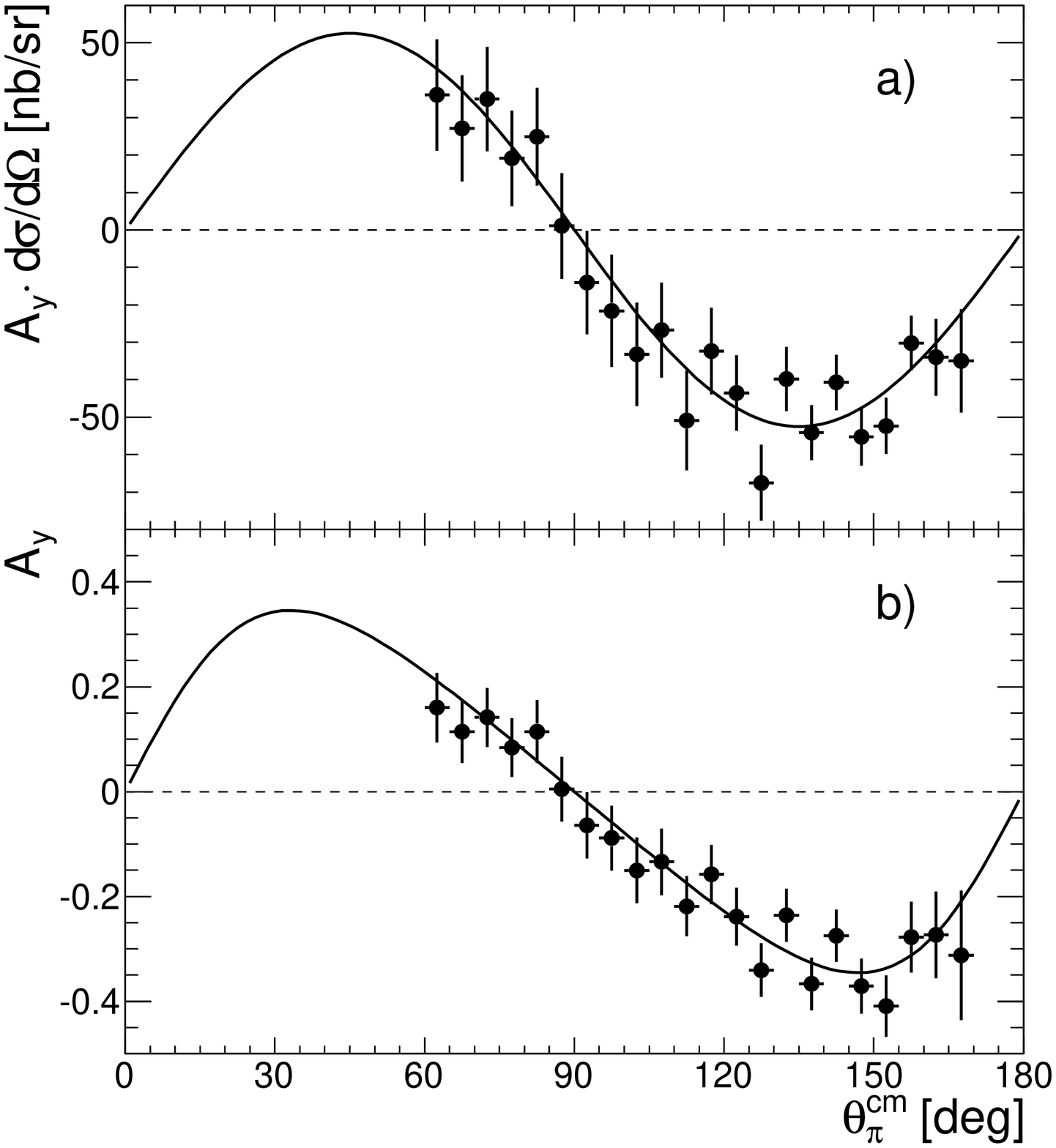}
\caption{\label{gig4} (a) The product of the measured analysing power and
differential cross section for the \vpppi\ reaction, with cross section
errors as in Fig.~\ref{gig3}. The curve represents the best fit of
Eq.~(\ref{aydsig}), with $b_2=1.82~\mu$b/sr and all higher terms eliminated.
(b) Measured values of $A_y$ for the \vpppi\ reaction. The errors shown are
purely statistical; the overall systematic uncertainty is about 5\%. The line
represents the quotient of the best fit in panel-a and the fit to the cross
section in Fig~\ref{gig3}.}
\end{figure}

The results for the analysing power of the \vpppi\ reaction are displayed in
Fig.~\ref{gig4}, with $A_y(d\sigma/d\Omega)$ being shown in panel a and $A_y$
in panel b. These observables must be antisymmetric about $90^\circ$ and the
crossing of the data through zero around this angle is some confirmation of
our estimation of $L_\uparrow/L_\downarrow$. These data are subject to the
overall uncertainties associated with the luminosity and acceptance
evaluation, though these are not relevant for the $A_y$ in Fig.~\ref{gig4}b.
There remains, however, the $\pm3\%$ arising from the uncertainty in the
value of $L_\uparrow/L_\downarrow$.

The $A_y\,d\sigma/d\Omega$ data are consistent with a
$\sin\theta_{\pi}\cos\theta_{\pi}$ behaviour and a fit using the general form
of Eq.~(\ref{aydsig}) yields
\begin{equation}
\label{ayval} b_2=1.82\pm0.10~\mu\textrm{b/sr}
\end{equation}
with $\chi^2/\textrm{NDF} = 15/21$. There is therefore no evidence for any
$\sin\theta_{\pi}\cos^3\theta_{\pi}$ dependence, i.e., a non-zero $b_4$
coefficient. The resulting fits are shown in Fig.~\ref{gig4}.

In order to understand the significance of the results reported here, we must
attempt a partial wave description of the data. The most general form of the
reaction amplitude is
\begin{equation}
\label{amps} \mathcal{M}=A\boldsymbol{S}\cdot\hat{\boldsymbol{p}}+B\boldsymbol{S}\cdot\hat{\boldsymbol{k}},
\end{equation}
where $\boldsymbol{S}$ is the polarisation vector of the initial $pp$
spin-triplet state. $\hat{\boldsymbol{p}}$ and $\hat{\boldsymbol{k}}$ are
unit vectors in the c.m.\ frame along the directions of the incident proton
and final pion, respectively.

The observables studied here are expressed in terms of the two scalar
amplitudes $A$ and $B$ through~\cite{HAN2004}
\begin{eqnarray}
\nonumber\left(\frac{d\sigma}{d\Omega}\right)_{\!0}&=&\frac{k}{4p}\left(|A|^2+|B|^2
+2\,\textrm{Re}[AB^*]\cos\theta_{\pi}\right)\!,\\
\label{obs1}
A_y\left(\frac{d\sigma}{d\Omega}\right)_{\!0}&=&\frac{k}{4p}\left(
2\,\textrm{Im}[AB^*]\sin\theta_{\pi}\right)\!.
\end{eqnarray}

The experimental data show no evidence for high partial waves
at 353~MeV and so we model these results with only $\ell=0$ and
$\ell=2$ contributions. The latter can arise from initial $L=1$
or $L=3$ waves so that, in total, there are three possible
transitions, $^{3\!}P_0\to\, ^{1\!}S_0s$, $^{3\!}P_2\to\,
^{1\!}S_0d$, and $^{3\!}F_2\to\, ^{1\!}S_0d$, see e.g.\
Ref.~\cite{PIA1986} for the explicit form of the spin-angular
structures. We denote the corresponding amplitudes by $M_s^P$,
$M_d^P$, and $M_d^F$, respectively.

Expanding the scalar amplitudes in terms of these partial waves gives
\begin{eqnarray}
\nonumber
A&=&M_s^P-\fmn{1}{3}M_d^P+M_d^F\left(\cos^2\theta_{\pi}-\fmn{1}{5}\right)\!,\\
\label{amps2} B&=&\left(M_d^P-\fmn{2}{5}M_d^F\right)\cos\theta_{\pi}.
\end{eqnarray}

Equations~(\ref{obs1}) and (\ref{amps2}) then allow one to relate the
measured observables of Eqs.~(\ref{dsig}) and (\ref{aydsig}) to the partial
wave amplitudes. For consistency, since we have neglected any possible
effects arising from $s$-$g$ interference, we shall also drop terms that are
bilinear in $d$-wave production amplitudes. In this approximation
\begin{eqnarray}
\nonumber
a_0&=&|M_s^P|^2-\fmn{2}{3}\textrm{Re}\left[M_s^P(M_d^P+\fmn{3}{5}M_d^F)^*\right]\!,\\
\nonumber
a_2&=&2\,\textrm{Re}\left[M_s^P(M_d^P+\fmn{3}{5}M_d^F)^*\right]\!,\\
\label{relation} b_2&=&
2\,\textrm{Im}\left[M_s^P(M_d^P-\fmn{2}{5}M_d^F)^*\right]\!,
\end{eqnarray}
and so the data only provide three relations between the three
complex amplitudes. The transverse spin correlation parameters
contain no extra information since $A_{y,y}=1$ and this is also
true for $A_{x,x}$ up to $d$-$d$ interference terms. If the
longitudinal-transverse spin correlation parameter $A_{x,z}$
were measured, this would provide one further relation but this
would still not be sufficient for an unambiguous partial wave
decomposition. For this we need information about the phases of
the production amplitudes.

The $^{3\!}P_0$ partial wave is uncoupled and, at the energy where the
experiment was performed, its inelasticity is very small. Under these
conditions the Watson theorem, which fixes the phase induced by the initial
state interaction to that of the elastic proton-proton scattering,
applies~\cite{MW}. Thus we take $M_s^P=|M_s^P|e^{i\delta_{^{3\!}P_0}}$, with
$\delta_{^{3\!}P_0}=-14.8^{\circ}$~\cite{SAID}. Note that we do not include
any phase associated with the $^{1\!}S_0$ final $pp$ state because it is
common for all partial waves and therefore does not affect the observables.

For coupled channels, such as $^{3\!}P_2-\,^{3\!}F_2$, the strict conditions
of the Watson theorem do not apply. However, at our energy the mixing
parameter, as well as the inelasticities, are still negligibly small and thus
to a good approximation we may also use the Watson theorem here. Further
evidence for the smallness of the channel coupling is to be found in two
potential models~\cite{Mac1987,Hai1993}. In both models the $T$-matrix for
the transition from the $^{3\!}F_2$ to the $^{3\!}P_2$ wave is almost real;
the phase of $M_d^P$ is driven by $\delta_{^{3\!}P_2}=17.9^{\circ}$, whereas
the phase of $M_d^F$ can be neglected. The quality of this approximation was
also checked by explicit calculations of the $d$-wave production amplitudes
within chiral effective field theory. These showed that up to order
$m_{\pi}/m_N$ the above phase assumptions should be valid to within $\pm
2^{\circ}$.

Using the phase information in this way, we find that
\begin{eqnarray}
\nonumber M_s^P&=&\phantom{-}(55.3\pm0.4)-(14.7\pm0.1)i~\sqrt{\textrm{nb/sr}}\,,\\
\nonumber M_d^P&=&-(26.6\pm1.1)-(8.6\pm0.4)i~\sqrt{\textrm{nb/sr}}\,,\\
\label{result} M_d^F&=&\phantom{-}(5.3\pm2.3)~\sqrt{\textrm{nb/sr}}\,.
\end{eqnarray}
The values quoted here were obtained by considering also our \nppppi\ data
though the numbers would change but marginally if one included only the
\pppi\ results in the fit. The error bars quoted here are statistical and do
not include the overall systematic uncertainties. However, changing the
normalisations of the differential cross section and analysing powers by 3\%
and 4\%, respectively, leads to changes that are comparable to the quoted
errors. On the other hand, we could not investigate the less tangible ones
associated with the neglect of the channel coupling and the truncation in the
partial wave expansion. The weakness of pion production from the initial
$^{3\!}F_2$ waves at 353~MeV, in addition to being in agreement with
theoretical prejudices, is also consistent with the low inelasticity found
for this wave~\cite{SAID}.

In summary, we have measured the differential cross section and analysing
power of the \vpppi\ reaction at 353~MeV. The angular distributions of $A_y$
and $d\sigma/d\Omega$ are both well represented by retaining only pion $s$
and $d$ waves in a phenomenological description. The values of
$d\sigma/d\Omega$ agree well with the results obtained at
CELSIUS~\cite{BIL2001} over most of the angular range. However, at this
energy these data flatten off around the middle of the angular distribution
and, if this effect were correct, it would signal a large contribution from
$^{3\!}F_2$ or even higher partial waves.

By making plausible assumptions on the coupling between the nucleon-nucleon
channels and invoking the Watson theorem it was possible to estimate the
partial wave amplitudes with their phases. These could be checked through a
future measurement of the spin-correlation parameter $A_{x,z}$, though this
would require the installation of a Siberian snake in COSY.

In an associated letter~\cite{DYM2011}, the isospin-0 amplitudes are
investigated through the measurements of the differential cross section and
analysing power of the quasi-free $\pol{p}n\to \{pp\}_s\pi^-$ reaction in
this energy domain. The extraction of the $p$-wave amplitudes from the data
of Ref.~\cite{DYM2011} required a knowledge of the $s$- and $d$-wave
amplitudes of the type provided here. In addition, data have already been
taken on the transverse spin correlation parameter $A_{x,x}$ for this
reaction~\cite{DYM2010}. The full collection of these results will lead to
very useful constraints on the parameters of the chiral effective field
theory that link pion production to the three-nucleon
interaction~\cite{HAN2000,EPE2002}.

We are grateful to other members of the ANKE Collaboration for
their help with this experiment and to the COSY crew for
providing such good working conditions, especially of the
polarised beam. This work has been partially supported by the
BMBF (grant ANKE COSY-JINR), RFBR (09-02-91332), DFG (436 RUS
113/965/0-1), the JCHP FFE, the SRNSF (09-1024-4-200), the
Helmholtz Association (VH-VI-231), STFC (ST/F012047/1 and
ST/J000159/1), and the EU Hadron Physics 2 project ``Study of
strongly interacting matter''
%
%

%
%
\end{document}